\documentclass
[superscriptaddress,secnumarabic,amssymb,amsmath,nobibnotes,aps,prd,showkeys,showpacs,twocolumn,nofootinbib]{revtex4-1}%

\usepackage{bm}
\usepackage{mathrsfs}
\usepackage[all]{xy}
\usepackage{epsfig,subfigure}
\usepackage{latexsym, amssymb, amsmath, amsfonts}
\usepackage[english]{babel}
\usepackage[OT1]{fontenc}
\usepackage[utf8]{inputenc}
\usepackage{makeidx}
\usepackage[colorlinks=true,breaklinks=true]{hyperref}
\usepackage{graphicx, wrapfig, epsf}
\usepackage{float}
\usepackage[dvipsnames, table]{xcolor}
\hypersetup{urlcolor=BlueViolet,
	        citecolor=Plum,
	        linkcolor=OliveGreen}
 \usepackage{url}

\definecolor{Gray}{rgb}{.9,.9,.9}

\newcommand{\apjl}{{The~Astrophysical~Journal~Letters}}

\newcommand{\mnras}{{MNRAS}}

\begin{document}

\title{Fermi gas and modified gravity}

\author{Aneta Wojnar}
\email[E-mail: ]{awojnar@ucm.es}
\affiliation{Laboratory of Theoretical Physics, Institute of Physics, University of Tartu,
W. Ostwaldi 1, 50411 Tartu, Estonia
}
\affiliation{Departamento de F\'isica Te\'orica, Universidad Complutense de Madrid, E-28040, Madrid, Spain}

\begin{abstract}
We derive the equations of state of Fermi gas by maximizing the Fermi-Dirac entropy in modified gravity in relativistic and non-relativistic case. It will be demonstrated that the microphysics must depend on a given theory of gravity in order to consistently describe a physical system at a statistical equilibrium state.
\end{abstract}

\maketitle

\section{Introduction}

Since 1998, the need for an alternative to General Relativity (GR) has been still growing as the fact of the accelerated expansion of the Universe \cite{huterer1999prospects} has not been yet explained in a satisfactory way. Many of those approaches also try to address puzzles such as mentioned dark energy, believed to be an agent of the accelerated expansion, as well as dark matter \cite{copeland2006dynamics,nojiri2007introduction,NOJIRI201159,nojiri2017modified,Capozziello:2007ec,PhysRevD.71.063513} (for review, see \cite{saridakis2021modified}), or to understand the problem
of spacetime singularities \cite{Senovilla:2014gza}. There are also attempts to unify physics of different scales \cite{parker2009quantum,birrell1984quantum} as well as to propose a model explaining the existence of massive compact objects whose masses exceed theoretical predictions \cite{2018ApJ...859...54L,2013Sci...340..448A,2020ApJ...896L..44A,2020PhRvL.125j1102A,2020PhRvL.125z1105S}. On the other hand, multimessenger astronomy \cite{abbott2017gw170817,abbott2017gravitational,2018FrASS...5...44E,2022MNRAS.513.4159P,abbott2019tests} is a new tool allowing to test theories of gravity, making that studying compact objects and their matter properties is a very up-to-date topic. Moreover, missions such as already lunched GAIA \cite{gaia} and James Webb Space Telescope \cite{webb}, or Nancy Grace Roman Space Telescope \cite{roman}
will provide more accurate data on stellar and substellar objects, such that studying those bodies in alternative to GR theories of gravity is another possibility to constrain modified gravity.
So far, a few such tests have been proposed with the use of stellar and substellar objects \cite{saito2015modified,kozak2021invariant,olmo2021parameterized} (for review, see \cite{olmo2020stellar}). The most common are altered limited masses, for example, the Chandrasekhar mass of white dwarfs \cite{chandrasekhar1935highly,Saltas2018white,jain2016white,banerjee2017constraints,wojnar2021white,belfaqih2021white,sarmah2022stability,kalita2022weak}, the minimum Main Sequence mass \cite{sakstein2015hydrogen,sakstein2015testing,crisostomi2019vainshtein,Olmo:2019qsj,Kalita:2021zjg}, Jeans \cite{capozziello2012jeans} and opacity mass \cite{wojnar2021jupiter}, or minimum mass for deuterium burning \cite{rosyadi2019brown}. Modified gravity can also alter the light elements' abundances in stellar atmospheres \cite{Wojnar:2020frr}. The Sun turn out to be also a promising laboratory to test theories of gravity by the mean of helioseismology \cite{saltas2019obtaining,saltas2022searching}. The evolution of non-relativistic stars \cite{Wojnar:2020txr, chowdhury2021modified,Guerrero:2021fnz, straight2020modified,Gomes:2022sft} as well as brown dwarfs \cite{Benito:2021ywe,Kozak:2022hdy} and planets \cite{wojnar2021jupiter,Wojnar:2022ttc}, as well as their internal properties \cite{Kozak:2021ghd,Kozak:2021zva,Kozak:2021fjy} can also be tools to test gravitational proposals. 

However, let us notice that the common approach followed in the mentioned literature is mainly related to obtaining some observables in a given model of gravity when {\it only} gravitational part of the structural equations describing stellar and substellar objects is modified. Energy generation ratios, opacities, and equations of state (EoS's) are assumed to be the ones which are used in Newtonian gravity or GR (but notice that in order to obtain values of some microscopic variables one sometimes uses the solutions of modified hydrostatic equilibrium equations). The question we would like to answer in this paper is if we can really follow that approach to describe physical phenomena. 

The reasons why one wonders if the set of equations we are using in modified gravity is consistent are the indications which have appeared in the previous studies, such as showing that chemical potential depends on gravity \cite{kulikov1995low}, therefore any change in the description of the gravitational field should also affect it. It was also demonstrated that modified gravity changes the geodesic deviation equation on the star's surface taking the form of the Hook's law, and consequently, introducing corrections to the polytropic equation of state \cite{kim2014physics}. Moreover, modifications to the gravitational interactions are also manifested in microscopic quantities, such as for example opacity, suggesting to treat them as effective ones \cite{sakstein2015testing}. Thermonuclear processes happening in the interiors of stars may also have different theoretical description when modified gravity is taken into account, following the fact that one needs a theory of gravity to compute the energy generation rate \cite{sakstein2015hydrogen,Olmo:2019qsj,crisostomi2019vainshtein,rosyadi2019brown,wojnar2021lithium}. Since some theories of gravity can introduce a dependence of the metric on the local energy-momentum distributions, the elementary particle interactions are also affected by these gravitational proposals \cite{delhom2018observable}. Furthermore, it was shown that specific heats of electrons and ions, Debye temperature and crystallization processes in white dwarfs are also dependent on a model of gravity \cite{2022arXiv221204918K}. An interesting fact that chemical reactions rates depend on gravity \cite{lecca2021effects} also suggests that any modification to this interaction will also have an effect on the kinetic rate constants of chemical reactions. Apart from this, ignoring the relativistic effects introduced by GR causes underestimation of the limiting masses of compact stars - but computations of equations of state in curved spacetime provide that in the degenerated stars one deals with EoS which depend explicitly on the metric components \cite{hossain2021equation,hossain2021higher}, therefore one should expect similar changes in EoS derived from different assumptions on the gravitational interaction. It is mainly manifested by already mentioned enhanced chemical potentials and temperatures \cite{li2022we} which are a part of the equations' set 
describing a statistical system.

One also alters thermodynamics quantities and EoS's when dealing with (pseudo-)scalar fields, for instance, axions \cite{sakstein2022axion}.
Another manifestation of special relativity and gravity appearing as the corrections to the EoS's and other microscopic variables is provided by the Generalized Uncertainty Principles which is, roughly speaking, a proposal of how the dispersion relation between energy, mass and momentum represented by the speed of light $c$, and gravity, given by the Newtonian constant $G$, should be present in Heisenberg's Uncertainty principle \cite{moussa2015effect,rashidi2016generalized,belfaqih2021white,mathew2021existence,hamil2021new,gregoris2022chadrasekhar}. 

In what follows, we will focus on the derivation of the Fermi gas EoS by the maximization of the Fermi entropy in modified gravity. Although it can be done for a more general system, Fermi gas is a model of matter with an exceptional relevance in physics of stars and substellar objects. It is widely used to describe some parts of the interiors of neutron and white dwarf stars, as well as non-relativistic stars, such as, for instance, pre- and Main Sequence ones, or substellar objects, that is, brown dwarfs and planets. Let us notice that there were some problems reported in the case of polytropic stars\footnote{Polytropic equation of state is a special case of the Fermi gas, as we will see further.} in Palatini-like gravities \cite{barausse2008no,barausse2008curvature}, however, further studies have shown that it is not the case \cite{kim2014physics,olmo2008reexamination,Wojnar:2018hal}. It is so because of the wrong assumptions - the region close to the surface of the star is not modelled by the polytrope - as well as computational errors \cite{kim2014physics,olmo2008reexamination}. Moreover, performing correctly the matching between the interior and exterior solutions also showed that the potentially pathological effects are shifted beyond the domain of physical interest \cite{2020CQGra..37u5002O}. The numerical analysis also do not reported any singular behaviour \cite{Wojnar:2018hal}. Different forms of the polytropic equations of state were also used in case of non-relativistic objects \cite{Benito:2021ywe,Kozak:2021ghd}, in which again any problems appeared.

To derive the Fermi gas EoS, we will firstly introduce one of the simplest generalization of GR provided by Palatini $f(\hat R)$ gravity in the section \ref{palatini}. We will focus on the spherical-symmetric objects and we will provide the exact solutions of the field equations for an arbitrary $f(\hat R)$. In the section \ref{nrl} we will briefly discuss the non-relativistic limit of the theory, followed further by recalling the local and global variables of a system of self-gravitating fermions restricted to a box with a finite radius. We will also calculate the most probable state of such an isolated system providing a set of equations which describe it. Among them, the modified hydrostatic equilibrium as well as modified EoS's can be found. The relativistic effects introduced to the analogous procedure are discussed in the section \ref{rc}. Our summary and conclusions are given in the section \ref{conl}. 
We use $\kappa^2=8\pi G/c^4$ and the $(-+++)$ signature convention.

\section{Palatini gravity}\label{palatini}

Palatini $f(\hat R)$ gravity is one of the simplest case of the so-called metric-affine models of gravity. In such an approach, one gives up the assumption on the Levi-Civita connection, that is, that the connection $\hat \Gamma$ is compatible with the physical metric appearing in the gravitational as well as matter part of the action. Therefore, we are dealing with two independent geometric structures, with the $f(R)$ gravity-like action:
\begin{equation}
    S[g,\hat\Gamma,\psi_m]=\frac{1}{2\kappa^2}\int \sqrt{-g}f(\hat R) d^4 x+S_{\text{matter}}[g,\psi_m],\label{action}
\end{equation}
where the Palatini curvature scalar is given by $\hat R = \hat R_{\mu\nu}(\hat\Gamma)g^{\mu\nu}$. It is a well-known fact that the Ricci-Palatini scalar must be symmetric, since its antisymmetric part introduces instabilities \cite{Borowiec:1996kg,Allemandi:2004wn,BeltranJimenez:2019acz}. Let us notice that the connection $\hat\Gamma$ is not coupled to the matter fields $\Psi_m$, therefore particles moves along the geodesics indicated by the Levi-Civita connection $\Gamma$ of the metric $g$. As already mentioned, in general, $\hat\Gamma \neq \Gamma$.

Since we are dealing with two independent objects, the variation of the above action is taken with respect to the metric $g$ and the connection $\hat\Gamma$. The first procedure gives
\begin{equation}
f'(\hat{R})\hat{R}_{\mu\nu}-\frac{1}{2}f(\hat{R})g_{\mu\nu}=\kappa^2 T_{\mu\nu},\label{structural},
\end{equation}
where the energy-momentum tensor  $T_{\mu\nu}=-\frac{2}{\sqrt{-g}}\frac{\delta S_m}{\delta g_{\mu\nu}}$ and prime denotes in this particular case\footnote{In further part of the paper, the prime will denote differentiating with respect to the radial coordinate $r$; however, it will be always clear what is the notation we are currently using.} differentiating with respect to the curvature $\hat{R}$. In the further part of the paper, we will use the perfect fluid energy-momentum tensor which has the follwoing form:
\begin{equation}
    {T}_{\mu\nu}=\left({\epsilon}+\frac{P}{c^2}\right){u}_\mu {u}_\nu +{P} g_{\mu\nu},
\end{equation}
where ${u}^\mu$ is a four-velocity vector field co-moving with the fluid, with the normalization condition $g_{\mu\nu}u^\mu u^\nu =-c^2$.

A very useful equation, which is the result of the contraction of \eqref{structural} with the inverse of the metric $g_{\mu\nu}$, is 
\begin{equation}\label{trace}
    f'(\hat{R})\hat{R}-2f(\hat{R})=\kappa^2 \mathcal{T},
\end{equation}
giving the algebraic relation between the scalar curvature and the trace of the energy momentum tensor $\mathcal{T}=g^{\mu\nu}T_{\mu\nu}$ for a chosen $f(\hat R)$. It is a generalization of the GR case, in which the trace of the field equations provides $ R:=R_{\mu\nu}(\Gamma(g)) g^{\mu\nu} = -\kappa \mathcal{T}$. Therefore, Palatini field equations can be interpreted as GR ones with modified matter part since all modifications turn out to be functions of baryonic matter fields and their derivatives\footnote{That is, one can rewrite the field equation \eqref{structural} as $G_{\mu\nu}(g)=\kappa T^\text{eff}_{\mu\nu}$, to see details in, for instance, \cite{2011IJMPD..20..413O}.}.

On the other hand, the variation with respect to the independent connection provides, after some algebraic transformations, 
\begin{equation}
\hat\nabla_\beta(\sqrt{-g}f'(\hat{R}(T))g^{\mu\nu})=0,\label{con}
\end{equation}
where the covariant derivative is ruled by the independent connection $\hat\Gamma$. If we introduce a metric $h$ which is conformal to the metric $g$ in the following way
\begin{equation}
    h_{\mu\nu} = f'(\hat{R}(T))g_{\mu\nu},
\end{equation}
it turns out that the equation for the connection \eqref{con} is nothing else but 
\begin{equation}
\hat\nabla_\beta(\sqrt{-h}h^{\mu\nu})=0,\label{con2}
\end{equation}
that is, $\hat\Gamma$ is the Levi-Civita connection of the metric $h$. The existence of the conformal structures in this model of gravity allows us to introduce the "Einstein frame", in which the equations have easier forms, and hence it can simplify rather tedious calculations \cite{afonso2018mapping,Olmo:2021yul}. It can be shown that after performing the conformal transformation one ends up with the field equations given as
 \begin{subequations}
	\begin{align}
	\label{EOM_P1}
	 \bar R_{\mu\nu} - \frac{1}{2} h_{\mu\nu} \bar R  &  =\kappa \bar T_{\mu\nu}-\frac{1}{2} h_{\mu\nu} \bar U(\phi)
	\end{align}
	\begin{align}
	\label{EOM_scalar_field_P1}
	  \phi\bar R &  -  (\phi^2\,\bar U(\phi))^\prime =0,
	\end{align}
\end{subequations}
where prime in the above equation denotes the derivative with respect to the non-dynamical scalar field $\phi$ defined as $\phi=f'(\hat{R})$, with its the potential $\bar U(\phi)=\frac{\hat{R}\phi-f(\hat{R})}{\phi^2}$. Let us notice that the energy-momentum tensor in the Einstein frame is related to the physical one by $\bar T_{\mu\nu}=\phi^{-1}T_{\mu\nu}$.

As already mentioned, it is easier to work in Einstein frame. For example, using the procedure demonstrated in \cite{Wojnar:2016bzk,Wojnar:2017tmy}, we can write, in the case of the perfect fluid, the interior solutions of the above modified field equations in the case of the spherical-symmetric spacetime
\begin{equation}
    d\hat s^2=-\hat B(\hat r)dt^2 + \hat A(\hat r)d\hat r^2 +\hat r^2d\Omega^2
\end{equation}
straight away as
\begin{align}
    \hat A(\hat r) &= \left( 1-\frac{2GM(\hat r)}{\hat r} \right)^{-1},\\
    \hat B(\hat r) &= \exp{\left[-\int^\infty_{\hat{r}} \frac{2G}{\tilde r^2}\Big(M(\tilde r)-4\pi \tilde r^3 \hat \Pi(\tilde r)\Big)\right]\hat A d\tilde r},
\end{align}
where the mass function is defined in the following way
\begin{equation}\label{mass0}
    M(\hat r)=\frac{1}{c^2}\int 4\pi'\tilde r^2 \hat Q(\tilde r) d\tilde r
\end{equation}
while the generalized density and pressure in the Einstein frame are, respectively,
\begin{align}
\hat{Q} & = \hat{\epsilon} + \frac{\bar U(\phi)}{2\kappa^2c^2},\label{genden} \\
\hat{\Pi} & = \hat{P} - \frac{\bar U(\phi)}{2\kappa^2}, \label{genpres}
\end{align}
with $\hat{\epsilon}= \epsilon/\phi^2$ and $\hat{P}= P/\phi^2$.
This procedure also allows to write down the Tolman-Oppenheimer-Volkoff (TOV) equation in the familiar form
\begin{equation}
     \frac{d\hat \Pi}{d\hat r}=-\frac{\hat Q(\hat r)+\hat \Pi(\hat r)}{c^2}\frac{\frac{GM(\hat r)}{\hat r^2}+\frac{4\pi G}{c^2}\hat \Pi \bar r}{1-\frac{2GM(\hat r)}{\hat rc^2}}.
\end{equation}
Let us recall that in order to consider a physical spherical-symmetric system, one must transfer the above equations back to the Jordan frame. This provides the metric components of $g_{\mu\nu}$ as ($\hat r^2 = \phi r^2$)
\begin{align}
    B(r) =& \phi^{-1}\text{exp}\left[-\int_r^\infty \frac{2G}{\phi^\frac{1}{2} \tilde r^2} \left(M(\tilde r)\right.\right.\label{sol1}\\
     -&\left.\left. 4\pi \tilde r^3 \phi^{-\frac{1}{2}}\Big(P(\tilde r)-\frac{U(\phi)}{2\kappa}\Big)\right) \left( 1-\frac{\tilde r}{2}\partial_{\tilde r} ln\phi\right)A(\tilde r)d\tilde r  \right],\nonumber\\
      A(r) =& \left( 1-\frac{ r}{2}\partial_{ r} ln\phi\right)^2 \left( 1-\frac{2GM( r)}{\phi^{1/2} r}\right)^{-1}
\end{align}
where $U(\phi)=\hat{R}\phi-f(\hat{R})$,
     while the mass function 
     \begin{equation}\label{mass}
         M=\frac{1}{c^2}\int^r_0 4\pi r^2 \frac{\epsilon(\tilde r) +\frac{U(\phi)}{2c^2\kappa}}{\sqrt{\phi}}\left( 1-\frac{\tilde r}{2}\partial_{\tilde r} ln\phi\right)^{-1}d\tilde r.
     \end{equation}
       The variable $\phi=\phi(r)$ is a function of the trace of the energy momentum tensor $\mathcal{T}$, that is, it is a function of the energy density $\epsilon(r)$ and pressure $ P(r)$. We will need those solutions in the further part of the paper.
     
    Although the presented work is for the general form of the Lagrangian $f(\hat R)$, let us briefly discuss the bounds of the theory's parameter, given by other works. In the case of the quadratic Lagrangian $f(\hat R)=\hat R+a \hat R^2$, the analytical consideration in the weak field limit provided that $|a| \lesssim 2\times 10^{12}\rm\,cm^2$ \cite{olmo2005gravity} while when one considers that electric and Newtonian gravitational forces are of the same order of magnitude, gives $|a| \lesssim 2\times 10^{9}\rm\,cm^2$ \cite{avelino2012eddington,jimenez2018born}. On the other hand, further studies shown that Solar System experiments cannot put the bound on the theory parameter because of the microphysics uncertainties \cite{Toniato:2019rrd}, while analysis with the SPARC catalogue data demonstrated that Palatini gravity, similarly as GR, also requires some amount of dark matter in order to explain the galaxy rotation curves \cite{Hernandez-Arboleda:2022rim}.

\section{Non-relativistic limit}\label{nrl}
Let us start with the non-relativistic case. It was demonstrated that Palatini $f(\hat R)$ gravity introduces additional terms to the Newtonian equations \cite{Wojnar:2018hal,Olmo:2021yul}, which, as already discussed, are proportional to the matter fields. Therefore, in vacuum, one deals with the usual Newtonian gravity. However, we are now focused on interiors of stellar and substellar objects, in which one cannot neglect the presence of matter. Going back to the details, we recall that for an analytic function \cite{Toniato:2019rrd} 
\begin{equation}\label{expan}
    f(\hat R)=\sum_{i=0}a_i\hat{R}^i
\end{equation}
the Poisson equation is given by \cite{Hernandez-Arboleda:2022rim}
\begin{equation}
    \nabla^2\Phi\approx \frac{\kappa^2}{2}(\rho +2a\nabla^2\rho)
\end{equation}
where $a\equiv a_2$ comes from the quadratic term of the Lagrangian (we neglect the cosmological constant, represented in \eqref{expan} by the parameter $a_0$) and $\rho$ is the mass density explained in more detailed below. Therefore, non-relativistic studies can only reveal some information such as parameter's bounds, properties of the studied objects, etc., only for the quadratic model, since further term enters the equations on the sixth order \cite{Toniato:2019rrd}. For a spherical-symmetric spacetime one can rewrite the modified Poisson equation as
\begin{equation}
    \frac{1}{r^2}\frac{d}{dr}\left( r^2\frac{d}{dr}\Big(\Phi(r)-a\kappa^2\rho(r)\Big) \right)=\frac{\kappa^2}{2}\rho(r).
\end{equation}
 Integrating the above equation provides
\begin{equation}
    \Phi'(r)=\frac{GM(r)}{r^2}+a\kappa^2\rho'(r),
\end{equation}
where $'$ denotes the derivative with respect to the $r$ coordinate. 
The mass function $M(r)$ in the non-relativistic limit is defined as usually\footnote{Let us notice that Palatini $f(\hat R)$ gravity does not introduce any extra degree of freedom as it may happen in more general proposals of metric-affine gravity \cite{d2022black,jimenez2022metric}.}
\begin{equation}
    M(r)=\int 4\pi'\tilde r^2 \rho(\tilde r) d\tilde r.
\end{equation}
If $r_s$ denotes the radius of a given astrophysical object, the boundary condition $M(r_s)=M$ with the assumptions $\rho(r_s)=0$ and $\rho'(r_s)=0$ results as
\begin{equation}
    \Phi(r)=-\frac{GM}{r}\;\;\textrm{for}\;\;r\geq R\;\;\;\;\textrm{and}\;\;\;\; \Phi(r_s)=-\frac{GM}{r_s},
\end{equation}
such that the gravitational potential at each point of spacetime can be expressed as
\begin{equation}\label{gravpot}
    \Phi(r)=-\frac{GM}{r}-4\pi G\int^R_r\Big( \rho(r)r-2a\rho'(r) \Big)dr.
\end{equation}
Finally, let us recall that the hydrostatic equilibrium equation has the well-known form,
\begin{equation}\label{hydro}
\frac{d\Phi}{dr}=-\rho^{-1}\frac{dP}{dr},
\end{equation}
which can be used in the modified Poisson equation
\begin{equation}\label{poisson}
    \frac{1}{r^2}\frac{d}{dr}
    \left(\frac{r^2}{\rho}\frac{dP}{dr}\right)=
    -4\pi G\left(\rho+\frac{2a}{r^2}\frac{d}{dr}\left[r^2\frac{d\rho}{dr}\right]\right).
\end{equation}
For a very special EoS, called polytrope, the above equation can be further transformed into the modified Lane-Emden equation, often used to study non-relativistic stellar and substellar objects \cite{Wojnar:2022txk}.

%%%%%%%%%%%%%%%%%%%%%%%%%%%%%%%%%%%%%%%%%5

\subsection{Fermi–Dirac entropy in a microcanonical ensemble}\label{nonrelF}

In what follows, we will follow the procedure excellently presented in \cite{chavanis2020statistical}. 

Before deriving the equation of state of Fermi gas and other microscopic properties of the system of self-gravitating fermions, let us recall the local and global variables describing such a system. The number density of fermions at position $  r$ with momentum $  p$ is given by $f(  r,   p)$d\textbf{r}d\textbf{p}, where $f(  r,   p)$ is the distribution function\footnote{Notice that it has nothing to the with the gravitational functional $f(\hat R)$ in \eqref{action}. It is common to use the letter $f$ for the distribution function and we will follow this notation, believing that it is always clear what object we are dealing with.}. Therefore, for an arbitrary distribution function the particle number density is given by
\begin{equation}\label{nu}
    n=\int f d  p
\end{equation}
while the kinetic energy density is defined as
\begin{equation}\label{ek1}
    \epsilon_\text{kin}=\int f \frac{p^2}{2m}d  p,
\end{equation}
where $E_\text{kin}=p^2/2m$ is the kinetic energy of a particle. Moreover, one expresses the local pressure in the form \cite{chandrasekhar1939book} 
\begin{equation}\label{presloc}
    P=\frac{1}{3}\int f\frac{p^2}{m}d  p,
\end{equation}
such that $P=\frac{2}{3}\epsilon_\text{kin}$. A combinatorial analysis taking into account the Pauli exclusion principle provides the Fermi-Dirac entropy density \cite{chavanis2004statistical,chavanis2006phase}
\begin{equation}\label{dent}
    s=-k_Bf_\text{max}\int \tilde f \,\text{ ln }\,\tilde f +
   (1- \tilde f) \,\text{ ln }\,(1-\tilde f)d  p,
\end{equation}
where $k_B$ is the Boltzmann constant, $f_\text{max}=\frac{g}{h^3}$ is the maximum value of the distribution function with $g=2s+1$ being the spin multiplicity of quantum states\footnote{$g=2$ for particles of spin $s=1/2$.}, and $h^3$ a volume of a microcell with no more than $g$ particles \cite{chavanis2020statistical}. We have defined $\tilde f= \frac{f}{f_\text{max}}$ for simplicity. It is important to comment that it can be demonstrated that the brief analysis presented here is valid for a general form of a entropy density although we focus only on the Fermi-Dirac one in this work.

Having defined the local variables, let us recall the global ones: the particle number $N$, mass $M$, energy $E$, and the Fermi–Dirac entropy $S$ are given as follows:
\begin{align}
    N=& \int n\, 4\pi r^2 dr,\label{pnum}\\
    M=&\, Nm=\int \rho\, 4\pi r^2 dr,\\
    E = &\, E^t_\text{kin} + W,\label{energy}\\
    S=&\,\int s \,4\pi r^2 dr,\label{entropy}
\end{align}
where the mass density is $\rho=nm,$ while $E^t_\text{kin}$ is the total kinetic energy given by
\begin{equation}
    E^t_\text{kin}=\int \epsilon_\text{kin} 4\pi r^2 dr= \int f\frac{p^2}{2m}4\pi r^2 drd  p.
\end{equation}
The gravitational potential energy  $W$, which is modified in Palatini gravity, is given as follows
\begin{equation}\label{epot}
    W=-\int \rho \textbf{r} \cdot \nabla\Phi d\textbf{r}= -\int \rho\left( 
\frac{GM}{r} + a\kappa r\rho'(r)    \right) dr,
\end{equation}
where the second equality is valid only for the spherical-symmetric case. 

Let us notice that for the systems in the microcanonical ensemble, the particle number $N$ and the energy $E$ are conserved. This fact will be used in the further part of the paper.

%%%%%%%%%%%%%%%%%%%%%%%%%

\subsection{Maximization of the Fermi–Dirac entropy}

Let us consider a system of particles, described by the equations given previously, restricted to a box of radius $r_s$ such that evaporation is prevented\footnote{One may study a canonical ensemble in which the thermal bath is allowed. Then, the procedure is quite analogous - one considers a minimization of the free energy $F=E-TS$ at fixed particle number $N$ instead.}. Then, for such a case, the most probable state of an isolated system is obtained by the maximization of the entropy (\ref{entropy}) at fixed energy (\ref{energy}) and particle number (\ref{pnum})
\begin{equation}
    \text{max}\, \{S\, |\, E, N\, \text{fixed}\},
\end{equation}
that is, one deals with a statistical equilibrium state in the microcanonical ensemble.

In order to obtained the exact expressions of the local and global variables, together with the Fermi EoS and hydrostatic equilibrium equation\footnote{This procedure will provide equations describing a system at a statistical equilibrium state.}, we will proceed as follows:
\begin{enumerate}
    \item We will obtain a local thermal equilibrium by maximizing the entropy density (\ref{dent}) at fixed kinetic energy $\epsilon_\text{kin}$ and particle number density $n(r)$ with respect to variations on $f$. This will allow to determine the distribution function $f$ and by this, the local variables.
    \item We will use the results obtained in the previous step to express the entropy $S$ as a function of the local variables and then we will  maximize it at fixed
energy $E$ and particle number $N$ with respect to variations on $\epsilon_\text{kin}$ and $n(r)$. Using the resulting laws of thermodynamics, we will show that an equation of state of the considered system depends on the considered modified gravity. It will turn out that such a dependence on a theory of gravity is also true for the other local variables describing the statistical equilibrium state of this system.
\end{enumerate}

\subsubsection{Local thermodynamic equilibrium}\label{nonreq}
 Maximizing the entropy density (\ref{dent}) at fixed kinetic energy $\epsilon_\text{kin}$ and particle number density $n(r)$ provides
 \begin{equation}\label{delta}
     \frac{\delta s}{k_B}-\beta(r)\delta\epsilon_\text{kin}+\alpha(r)\delta n =0,
 \end{equation}
 where $\alpha(r)$ and $\beta(r)$ are local, that is, position-dependent, Lagrangian multipliers. The variation of the considered entropy density with respect to the distribution function $f$ is given by
 \begin{equation}
     \delta s =-k_b \int \delta\tilde f \Big(\text{ ln }\tilde f-\text{ ln }\,(1-\tilde f)\Big)d  p.
 \end{equation}
Together with the variations of (\ref{nu}) and (\ref{ek1}), this leads to the Fermi-Dirac distribution function, having the following form
 \begin{equation}\label{dist}
     f(\textbf r, \textbf p)=\frac{g}{h^3}\frac{1}{1+\textrm{exp}\Big[ \frac{\beta(r)p^2}{2m}-\alpha(r) \Big]}.
 \end{equation}
It is the global maximum of entropy density at fixed $n$ and $\epsilon_\text{kin}$, that is, it is a condition of local thermodynamic equilibrium\footnote{It can be easily shown when the second variation of (\ref{delta}) is taken.}. The distribution function (\ref{dist}) can be written in a more familiar form
  \begin{equation}\label{dist2}
     f(\textbf r, \textbf p)=\frac{g}{h^3}\frac{1}{1+\textrm{exp}\Big[ \frac{p^2/2m-\mu(r)}{k_B T(r)} \Big]},
 \end{equation}
if we introduce the local temperature $T(r)$ and local chemical potential $\mu(r)$ by relating them with the multipliers:
 \begin{equation}
     T(r)=\frac{1}{k_B\beta(r)},\,\,\,\,\mu(r)=k_B T(r)\alpha(r).
 \end{equation}
 Therefore, the variational principle can be written as
 \begin{equation}\label{therm}
     ds=\frac{d\epsilon_\text{kin}}{T} - \frac{\mu}{T} dn,
 \end{equation}
 which is the well-known local first law of thermodynamics.
 
 On the other hand, using the distribution (\ref{dist2}) one can finally write explicitly the local variables (\ref{nu})-(\ref{dent}) as well as we can determine the temperature $T(r)$ and chemical potential $\mu(r)$ as functions of $n(r)$ and $\epsilon_\text{kin}$. Moreover, they also provide an equation of state as $P=P[n(r),T(r)]$. But more importantly for now, we can also derive the integrated Gibbs-Duhem relation from (\ref{dent})
 \begin{equation}\label{gibbs}
     s(r)=\frac{\epsilon_\text{kin}(r)+P(r)-\mu(r)n(r)}{T(r)},
 \end{equation}
 which will be used in the further steps. Let us also notice that the equation (\ref{therm}) can be rewritten in a more familiar form
 \begin{equation}\label{GD2}
     d\left(\frac{P}{T}\right)=n\,d\left(\frac{\mu}{T}\right)-\epsilon_\text{kin}d\left(\frac{1}{T}\right).
 \end{equation}
 
 %%%%%%%%%%%%%%%%%%%%%%%%%%%%%%%%%%%%%%
 \subsubsection{Variation of the entropy $S$}
 
 In order to maximize the entropy $S$ at fixed energy $E$ and particle number $N$
  \begin{equation}\label{delta2}
     \frac{\delta S}{k_B}-\beta_0\delta E+\alpha_0\delta N=0,
 \end{equation}
  where $\beta_0=\frac{1}{k_BT_0}$ and $\alpha_0=\frac{\mu_0}{k_BT_0}$ are global (uniform) Lagrange multipliers, respectively,
 we will use the integrate Gibbs-Duhem relation \eqref{gibbs} in (\ref{entropy}).

 With the use of (\ref{delta}) and the energy $E$ expressed as
 \begin{equation}
     E=\int \Big( \epsilon_\text{kin} + \frac{1}{2}nm \Phi \Big)dV,
 \end{equation}
 where $dV=4\pi r^2 dr$ is a volume element, the variational problem (\ref{delta2}) can be written as
 \begin{align}
     &\int \left( \frac{\delta \epsilon_\text{kin}}{k_BT}-\frac{\mu}{k_BT}\delta n \right)dV \\
     -\beta_0 &\int \left( \delta \epsilon_\text{kin} +m\Phi\delta n\right)dV +\alpha_0 \int \delta ndV=0,\nonumber
 \end{align}
 where the variations on $\delta \epsilon_\text{kin}$ and $\delta n$ must vanish identically. Therefore, vanishing of the first one provides that the temperature must be uniform, that is, $T=\text{const}$ at the statistical equilibrium. Because of that fact, the relation (\ref{GD2}) is now
 \begin{equation}\label{therm2}
     dP=nd\mu
 \end{equation}
such that using $\mu(r)=\mu[n(r),T]$ one finds
\begin{equation}
    \frac{dP}{dr}=n\frac{d\mu}{dr}.
\end{equation}
 On the other hand, vanishing of the variantion on the particle number density gives
 \begin{equation}\label{pot}
     \mu(r)=\mu_0 - m\Phi(r),
 \end{equation}
 where we have defined $\mu_0=\alpha_0 k_B T$. We also immediately notice the dependence on gravity. Taking the $r-$derivative of (\ref{pot}) results as
 \begin{equation}\label{pot2}
     \frac{d\mu}{dr}=-m\frac{d\Phi}{dr}=-m\left( \frac{GM(r)}{r^2}+a\kappa\rho'(r) \right),
 \end{equation}
 with the boundary condition $$\mu(r_s)=\mu_0-m\Phi(r_s)=\mu_0+\frac{GMm}{r_s}.$$
 It is clear that from (\ref{pot2}) and (\ref{therm2}) one gets the hydrostatic equilibrium equation (\ref{hydro}) for Palatini $f(\hat R)$ gravity in non-relativistic limit, that is,
 \begin{equation}\label{hydro2}
     \frac{dP}{dr}=-\rho(r)\left[ \frac{GM(r)}{r^2}+a\kappa\rho'(r) \right].
 \end{equation}
 We see that the modified hydrostatic equilibrium is implied by the condition of statistical equilibrium. It is so because of the chemical potential, which takes into account the presence of the gravitational field.

%%%%%%%%%%%%%%%%%%%%%%%%%%%%%%%%%%%%%%

\subsection{Statistical equilibrium state}
Let us now discuss forms of the local variables describing the statistical equilibrium. Using the results derived in the previous subsection and applying them into the distribution function (\ref{dist}) and equations (\ref{nu})-(\ref{presloc}) we have
\begin{align}
      f(\textbf r, \textbf p)=&\frac{g}{h^3}\frac{1}{1+e^{-\alpha_0}e^{\beta(p^2/2m+m\Phi(r))}},\\
      n(r)=&\frac{g}{h^3}\int\frac{d\textbf p}{1+e^{-\alpha_0}e^{\beta(p^2/2m+m\Phi(r))}},\label{loknu}\\
      \epsilon_{kin}(r)=&\frac{g}{h^3}\int\frac{p^2/2m\,d\textbf p}{1+e^{-\alpha_0}e^{\beta(p^2/2m+m\Phi(r))}},\\
      P(r)=&\frac{g}{3h^3}\int\frac{p^2/2m\,d\textbf p}{1+e^{-\alpha_0}e^{\beta(p^2/2m+m\Phi(r))}},\label{lokpres}
\end{align}
where $\beta=1/k_BT$ and $\alpha_0=\mu_0/k_BT$, while the gravitational potential is given by the equation (\ref{gravpot}). It can be demonstrated that one gets the hydrostatic equilibrium equation (\ref{hydro2}) by taking the $r-$derivative of (\ref{lokpres}) and applying to it the equation (\ref{loknu}). The above set of equations also provide the equation of state for self-gravitating non-relativistic Fermi gas, that is, using the properties of the Fermi integral $(u>0)$
\begin{equation}
    I_u(t)=\int^{+\infty}_0\frac{x^u}{1+te^x}dx,\,\;\;\,\,I'_u(t)=-\frac{u}{t}I_{u-1}(t),
\end{equation}
the equation of state $P=[\rho(r),T]$ at finite temperature can be written in the parametric form with the parameter $\alpha(r)=\alpha_0-\beta m\Phi(r)$ as
\begin{align}
    \rho(r)=&\frac{4\pi g\sqrt{2}m^{5/2}}{h^3\beta^{3/2}}I_{1/2}[e^{-\alpha_0+\beta m\Phi(r)}],\label{EoS1a}\\
     P(r)=&\frac{8\pi g\sqrt{2}m^{3/2}}{3h^3\beta^{5/2}}I_{1/2}[e^{-\alpha_0+\beta m\Phi(r)}]\label{EoS1b},
\end{align}
where the dependence on (modified) gravity is clearly evident, since $\Phi(r)$ is given by (\ref{gravpot}).

However, when one considers the complete degenerate Fermi gas at ground state, that is, $T\rightarrow0$, the Fermi-Dirac distribution \eqref{dist2} reduces to the Heaviside function
\begin{align}
    f(\textbf r, \textbf p)=&\frac{g}{h^3}\;\;\text{if}\;\; E_\text{kin}(p)<E_F(r) \\
     f(\textbf r, \textbf p)=&0\;\;\;\;\;\text{if}\;\; E_\text{kin}(p)>E_F(r),
\end{align}
where the Fermi energy $E_F$ depends on a model of gravity
\begin{equation}
    E_F(r)=\mu(r)=\mu_0 - m \Phi(r)
\end{equation}
as well as does the Fermi momentum, defined as
\begin{equation}\label{fermimom}
    p_F(r)=\sqrt{2m(\mu_0 - m \Phi(r))}.
\end{equation} 
The expression for the density and pressure simplify to
\begin{align}
    \rho=&\int f m d \mathbf p = \frac{4\pi g m}{3h^3}p_F^3(r),\\
    P=&\frac{1}{3}\int f \frac{p^2}{m} d\mathbf p = 
    \frac{4\pi g }{15mh^3}p_F^5(r),
\end{align}
which can be rewritten in a more familiar form, known as the polytropic EoS;
\begin{equation}
    P=K\rho^\frac{5}{3},\;\;\;
    K=\frac{1}{5}\left( \frac{3h^3}{4\pi g m^4} \right)^\frac{2}{3},
\end{equation}
without the direct dependence on gravity, in contrast to the Fermi momentum \eqref{fermimom}. Therefore, using the polytrope with the polytropic parameter $\gamma=5/3$ with the modified, non-relativistic hydrostatic equilibrium equation is consistent. However, when one deals with a more general form, as given by \eqref{EoS1a} and \eqref{EoS1b}, the effects of (modified) gravity should be taken into account.

%%%%%%%%%%%%%%%%%%%%%%%%%%%%%%%%%%%%%%%

\section{Relativistic case}\label{rc}
In this part, we will focus on the analogous formalism discussed in the previous section, but in the fully relativistic description of the gravitational field in modified gravity, as presented briefly in the section \ref{palatini}. We will recall again the local and global variables describing a system of self-gravitating fermions in the box of the radius $r_s$.

\subsection{Fermi–Dirac entropy in a microcanonical ensemble}

For an arbitrary distribution function $f(  r,   p)$ the particle number density is given by
\begin{equation}\label{nurel}
    n=\int f d  p
\end{equation}
while the energy density is defined as
\begin{equation}\label{edrel}
    \epsilon=\int f E(p)d  p,
\end{equation}
where $E$ is the total, that is, kinetic and rest mass, energy of a particle. It is given by
\begin{equation}
    E(p)=mc^2 + E_\text{kin}(p),
\end{equation}
where the relativistic kinetic energy 
\begin{equation}
    E_\text{kin}(p)=mc^2\left[ \sqrt{\frac{p^2}{m^2c^2}+1}-1 \right].
\end{equation} 
Moreover, the energy density \eqref{edrel} is expressed as
\begin{equation}
    \epsilon =\rho c^2 +\epsilon_\text{kin},
\end{equation}
with $\rho=nm$ being the rest mass density, and the kinetic energy density is defined by the following expression
\begin{equation}\label{ek}
    \epsilon_\text{kin}=\int f E_\text{kin}(p)d  p,
\end{equation}
while the local pressure
\begin{equation}\label{presloc2}
    P=\frac{1}{3}\int f\frac{dE(p)}{dp}d\textbf{p} =
    \frac{1}{3}\int f\frac{p^2 c^2}{\sqrt{p^2 c^2 + m^2 c^4}} d  p.
\end{equation}
Let us notice that the above expressions reduce to the forms in the nonrelativistic limit given in the section \ref{nonrelF} when $c\rightarrow +\infty$.

On the other hand, the Fermi-Dirac entropy density has the same form as before
\begin{equation}\label{dent2}
    s=-k_Bf_\text{max}\int \tilde f \,\text{ ln }\,\tilde f +
   (1- \tilde f) \,\text{ ln }\,(1-\tilde f)d  p.
\end{equation}

Regarding the global variables, the main difference with respect to the global variables defined in \ref{nonrelF} is that in the relativistic case we need to take into account the proper volume element. In order to significantly simplify the further calculations, we will use the "Einstein frame variables", which we denote with hat, that is, for instance, the radial coordinate in the Einstein frame is represented by $\hat r$, while in the physical frame it will be $r$. Therefore, the proper volume element in the Einstein frame is given by
$\hat \chi d\hat V$, where we have defined
\begin{equation}
    \hat \chi = : \left( 1-\frac{2GM(\hat r)}{\hat r} \right)^{-\frac{1}{2}},\quad d\hat V = :4 \pi \hat r^2 d\hat r.
\end{equation}
for further convenience. 

Therefore, the explicit forms of the entropy and particle number of the fermion gas are
\begin{align}
    S=&\,\int_0^{\hat r_s} s(\tilde r) \left( 1-\frac{2GM(\tilde r)}{\tilde r} \right)^{-\frac{1}{2}} \,4\pi \tilde r^2 d\tilde r,\label{entropy2}\\
     N=& \int_0^{\hat r_s} n(\tilde r)\, \left( 1-\frac{2GM(\tilde r)}{\tilde r} \right)^{-\frac{1}{2}}4\pi \tilde r^2 d\tilde r,\label{pnum2}
\end{align}
respectively, where $\hat r_s$ is the box's radius in the Einstein frame. The binding energy 
\begin{equation}
    E=\mathcal{E}-Nmc^2
\end{equation}
includes the rest mass energy $Nmc^2$, and the mass energy $\mathcal{E}=Mc^2$, with the mass $M$ given by \eqref{mass0}.

\subsection{Maximization of the Fermi–Dirac entropy in a microcanonical ensemble}

Local thermodynamic equilibrium state in the relativistic case is analogous to the non-relativistic one, as given in the section \ref{nonreq}. One has to maximaze the entropy density (\ref{dent2}) at fixed energy density $\hat\epsilon$ and particle number density $n$:
 \begin{equation}\label{deltarel}
     \frac{\delta s}{k_B}-\beta(\hat r)\delta\hat\epsilon+\alpha(\hat r)\delta n =0,
 \end{equation}
  where $\alpha(\hat r)$ and $\beta(\hat r)$ are local, Lagrangian multipliers in the Einstein frame.
Following the similar steps, the Fermi-Dirac distribution function is then
  \begin{equation}\label{distrel}
     f( \textbf r, \textbf p)=\frac{g}{h^3}\frac{1}{1+\textrm{exp}\Big[ \frac{E(p)-\mu(\hat r)}{k_B T(\hat r)} \Big]},
 \end{equation}
with the local first law of thermodynamics
\begin{equation}
         ds=\frac{d\hat\epsilon}{T} - \frac{\mu}{T} dn,
\end{equation}
and local variables\footnote{Notice that for the convenience, we use the hat only for the energy density and pressure since only those local variables appear in the solutions of field equations.}
\begin{align}
      n(\hat r)=&\frac{g}{h^3}\int\frac{d\textbf p}{1+\textrm{exp}\Big[ \frac{E(p)-\mu(\hat r)}{k_B T(\hat r)} \Big]},\\
     \hat \epsilon(\hat r)=&\frac{g}{h^3}\int\frac{E(p)d\textbf p}{1+\textrm{exp}\Big[ \frac{E(p)-\mu(\hat r)}{k_B T(\hat r)} \Big]},\\
       \epsilon_\text{kin}(\hat r)=&\frac{g}{h^3}\int\frac{E_\text{kin}(p)d\textbf p}{1+\textrm{exp}\Big[ \frac{E(p)-\mu(\hat r)}{k_B T(\hat r)} \Big]},\\
     \hat P(\hat r)=&\frac{g}{h^3}\int\frac{p\frac{dE(p)}{dp}d\textbf p}{1+\textrm{exp}\Big[ \frac{E(p)-\mu(\hat r)}{k_B T(\hat r)} \Big]}.
\end{align}
Using the above results with the Fermi-Dirac distribution function \eqref{distrel} in \eqref{dent2}, one finds the integrated Gibbs-Duhem relation 
\begin{align}
    s(\hat r)=\frac{\hat\epsilon(\hat r)+\hat P(\hat r)-\mu(\hat r)n(\hat r)}{T(\hat r)}
\end{align}
respectively. Let us notice that the integrated Gibbs-Duhem relation can be written in terms of the generalized density and pressure, given by \eqref{genden} and \eqref{genpres}, as
\begin{align}
    s(\hat r)=\frac{\hat Q(\hat r)+\hat \Pi(\hat r)-\mu(\hat r)n(\hat r)}{T(\hat r)},
\end{align}
such that the standard form of the first law of thermodynamics can be rewritten as 
\begin{equation}\label{1therm}
      d\left(\frac{\hat \Pi}{ T}\right)=n\,d\left(\frac{\mu}{ T}\right)-\hat Q d\left(\frac{1}{ T}\right).
\end{equation}

\subsection{Variation of the entropy $S$}
Using the above relations to the entropy definition \eqref{entropy2}
\begin{equation}
    S= \int^R_0 \frac{\tilde Q(\tilde r)+\tilde \Pi(\tilde r)-\mu(\tilde r)n(\tilde r)}{T(\tilde r)}  \hat \chi d\hat V
\end{equation}
and maximizing it at fixed mass energy $\mathcal{E}$ and particle number $N$
 \begin{equation}\label{delta3}
     \frac{\delta S}{k_B}-\beta_0\delta M+\alpha_0\delta N=0,
 \end{equation}
  where $\beta_0$ and $\alpha_0$ are global (uniform) Lagrange multipliers,
provides
\begin{align}\label{maxent}
    &\int \left( \frac{\delta \hat Q}{k_BT} -\frac{\mu}{k_B T} \delta n +\alpha_0 \delta n\right)\hat \chi d\hat V -\beta_0\int \delta\hat Q d\hat V\nonumber\\
    +&\int \left( \frac{\hat Q+\hat \Pi-\mu n}{k_BT} +\alpha_0 n \right)\delta\hat\chi d\hat V=0.
\end{align}
Moreover, we have the following useful expressions
\begin{align}
    \delta \hat \chi = \left( 1-\frac{2GM(\hat r)}{\hat r} \right)^{-\frac{3}{2}} \frac{G\delta M}{\hat r c^2}&,\label{id1}\\
    \delta M = \frac{1}{c^2}\int_0^{\hat r}\delta \hat Q 4\pi \tilde r^2 d\tilde r&,\\
    \frac{d\delta M}{d\hat r}=\frac{1}{c^2}\delta \hat Q 4\pi \hat r^2&.\label{id3}
\end{align}
Vanishing of \eqref{maxent} requires from the variations on $\delta n$ that the ratio between the local chemical potential and the local temperature is a constant value:
\begin{equation}\label{ratio}
    \alpha_0=\frac{\mu(r)}{k_BT(r)}\equiv \alpha.
\end{equation}
We will skip the index $0$ in the further part for simplicity. Using the above result in \eqref{1therm} provides that the first law of thermodynamics reduces to
\begin{equation}
 d\left(\frac{\hat \Pi}{ T}\right)=-\hat Q d\left(\frac{1}{ T}\right).
\end{equation}
It allows to write down the modified Tolman equation in the form
\begin{equation}\label{tolman}
\frac{d\hat \Pi}{d\hat r}=\frac{\hat Q(\hat r)+\hat \Pi(\hat r)}{T}\frac{dT}{d\hat r}.
\end{equation}
Inserting the constant ratio \eqref{ratio} in the above equation provides 
\begin{equation}
    \frac{d\hat \Pi}{d\hat r}=\frac{\hat Q(\hat r)+\hat \Pi(\hat r)}{\mu}\frac{d\mu}{d\hat r}.
\end{equation}
On the other hand, the variation on $\delta\hat Q$, after inserting \eqref{ratio} into \eqref{maxent}, gives
\begin{equation}
    \int \left[ \frac{\delta \hat Q}{k_BT}\hat \chi +  \left( \frac{\hat Q+\hat \Pi}{k_BT} \right)\delta\hat\chi  -\beta_0 \delta\hat Q \right]d\hat V =0.
\end{equation}
The above equation can be further rewritten with the use of \eqref{id1} and \eqref{id3} as
\begin{align}
    \int\left[ \left( \frac{\chi}{k_B T} - \beta_0\right)\frac{d\delta M}{d\hat r}
    +   \left( \frac{\hat Q+\hat \Pi}{c^2k_BT} \right)\frac{\partial\hat\chi}{\partial M}\delta M 4\pi \hat r^2 \right]d\hat r =0
\end{align}
while the integration by part applied to the first term will yield
\begin{align}
 0&=   c^2  \left( \frac{\hat\chi(\hat r_s)}{k_B T(\hat r_s)} - \beta_0\right)\delta M(\hat r_s)\\
   & -\int \left[c^2 \frac{d}{d\hat r}\left( \frac{\hat\chi}{k_B T} \right) - 
     \left( \frac{\hat Q+\hat \Pi}{k_BT} \right)\frac{\partial\hat\chi}{\partial M} 4\pi \hat r^2  \right] \delta M d\hat r. \nonumber
\end{align}
Vanishing of the first term in the above equation provides (we also skip the index $0$ in the Lagrangian multiplier $\beta_0$)
\begin{equation}
    \beta = \frac{\hat\chi(\hat r_s)}{k_B T(\hat r_s)} =  \frac{1}{k_B T(\hat r_s)\sqrt{1-\frac{2GM(\hat r_s)}{\hat r_s c^2}}},
\end{equation}
while vanishing of the second one
\begin{equation}
    c^2 \frac{d}{d\hat r}\left( \frac{\hat\chi}{ T} \right) = 
     \left( \frac{\hat Q+\hat\Pi}{T} \right)\frac{\partial\hat\chi}{\partial M} 4\pi \hat r^2  
\end{equation}
which can be expressed as 
\begin{equation}\label{tolman2}
    \frac{1}{T}\frac{d T}{d\hat r} = -\frac{1}{c^2}\frac{\frac{GM(\hat r)}{\hat r^2}+\frac{4\pi G}{c^2}\hat \Pi \hat r}{1-\frac{2GM(\hat r)}{\hat rc^2}}.
\end{equation}
This can be further rewritten in the well-known form of the Tolman-Oppenheimer-Volkoff (TOV) equation for Palatini gravity in the Einstein frame \cite{Wojnar:2017tmy} when we use the equation \eqref{tolman}
\begin{equation}
    \frac{d\hat \Pi(\hat r)}{d\hat r}=-\frac{\hat Q(\hat r)+\hat \Pi(\hat r)}{c^2}\frac{\frac{GM(\hat r)}{\hat r^2}+\frac{4\pi G}{c^2}\hat \Pi \hat r}{1-\frac{2GM(\hat r)}{\hat rc^2}},
\end{equation}
with the mass function given by
\begin{equation}
    M(\hat r) = \frac{1}{c^2}\int 4\pi \hat r^2 \hat Q d\hat r.
\end{equation}
Coming back to the physical frame, the TOV equation is simply the above one, after using the definitions \eqref{genden} and \eqref{genpres}, and performing the conformal transformation $\hat r^2 = \phi r^2$:
\begin{align}
     \frac{d P}{dr}=&-\frac{GM}{c^2r^2}\left(\epsilon+ P\right)\frac{1+\frac{4\pi r^3}{Mc^2} \left(P-\frac{U}{2\kappa^2}\right) }{1-\frac{2GM}{ \sqrt{\phi}rc^2}}\left(\phi-\frac{r\phi'}{2}\right) \nonumber \\
     &+\frac{4\kappa^2\left( P-\frac{U}{2\kappa^2} \right)\phi' + U_\phi \phi'\phi}{2\kappa^2\phi}
\end{align}
where $U_\phi=dU/d\phi$ while the mass function is given by the equation \eqref{mass}.

Apart from this, we are also able to write explicitly the Tolman and Klein equations, as well as the total particle number and the total entropy of the system. Combining equations \eqref{tolman} and \eqref{tolman2} and integrating the result with respect to $\hat r$, one can write the Tolman formula, which is a relation between the local temperature and the metric coefficient (here already written in the physical frame)
\begin{equation}
    T(r)=\,T_\infty  B(r)^{-\frac{1}{2}},
\end{equation}
where $T_\infty= T(r_s)\sqrt{1-\frac{2GM(r_s)}{r_s}}$ is the global temperature, that is, the temperature measured by the distant observer while $B(r)$ is given by the equation \eqref{sol1}. On the other hand, combining above equation with \eqref{ratio} provides the local chemical potential (Klein relation) as
\begin{equation}
     \mu(r)=\,\mu_\infty B(r)^{-\frac{1}{2}},
\end{equation}
where $\mu_\infty = \alpha k_BT(r_s)\sqrt{1-\frac{2GM(r_s)}{r_s}}$ is the global chemical potential, that is, the chemical potential measured by the distant observer (at infinity). Let us notice that both local variables depend on a theory of gravity via the interior solution, given in our case by \eqref{sol1}.
This is also a case of the total particle number and total entropy - after the conformal transformations, the equations \eqref{entropy2} and \eqref{pnum2} have the following form:
\begin{align*}
    N=& \int_0^R n( r)\, \left(1-\frac{2GM(r)}{\phi^\frac{1}{2}r}\right)^{-\frac{1}{2}}4\pi \phi(r) r^2 \nonumber\\
   &\times \left( 1-\frac{r}{2}\partial_r ln\phi(r) \right) d r,\\
        S=& \int_0^R \frac{\epsilon(r) +P(r)}{\phi(r)^\frac{3}{2}T(r)}\, \left(1-\frac{2GM(r)}{\phi^\frac{1}{2}r}\right)^{-\frac{1}{2}}4\pi  r^2 \nonumber \\
        &\times\left( 1-\frac{r}{2}\partial_r ln\phi(r) \right) d r -\frac{\mu_\infty}{T_\infty}N.
\end{align*}

In order to see directly how the Fermi equation of state is modified by Palatini gravity, let us introduce gravitational potential $\varphi(r)$:
\begin{equation*}
    B(r)=\left( \frac{\mu_\infty}{mc^2} \right)^2 \frac{1}{1+\frac{\varphi(r)}{c^2}},
\end{equation*}
such that the Fermi energy and momentum can be written as
\begin{equation*}
  E_F(r)=mc^2\sqrt{\frac{\varphi(r)}{c^2}+1},\;\; \;\;\;\;  p_F=m\sqrt{\varphi(r)}.
\end{equation*}
    Introducing $x=\frac{p_F}{mc}$ one writes the equation of state of the relativistic Fermi gas as
    \begin{align*}
          P &= \frac{\pi gm_\text{e}^4 c^5}{6 h^3}\left[x\left(2x^2-3\right)\sqrt{x^2+1}+3\sinh^{-1}x\right],\\
          \epsilon &= \frac{\pi gm_\text{e}^4 c^5}{2 h^3}\left[x\left(2x^2+1\right)\sqrt{x^2+1}-\sinh^{-1}x\right],\\
          n&=\frac{4\pi gm_\text{e}^3 c^3}{3 h^3}x^3
    \end{align*}                            
    which now clearly demonstrates that Fermi gas EoS depends on a model of gravity.

%%%%%%%%%%%%%%%%%%%%%%%%%%%%%%%%%%%%%%%%%%%%%%

\section{Conclusions}\label{conl}
Motivated by many indications suggesting that modified gravity can have something to say about the microscopic properties of stellar and substellar objects as briefly discussed in the introduction, we wanted to focus on that hypothesis. Therefore, 
in the following work we have analyzed the derivation of the Fermi equation of state in the non-relativistic and relativistic case, demonstrating that the effects of modified gravity indeed must be taken into account. However, before doing so, we have also shown how the local and global variables describing a system in the statistical equilibrium state depend on a theory of gravity. Therefore, the main conclusion of this paper is that the microphysics is dependent on the (modified) gravity.

In order to be consistent, one should take into account the effects of the considered model of gravity on microphysics. The common approach is to modify a hydrostatic equilibrium equation only, while other equations are kept as they resulted from GR or Newtonian physics. We have demonstrated in this paper that in order to deal with a consistent set of equations describing a stellar or substellar object in the statistical equilibrium state\footnote{And likely following by this, dynamical equilibrium state. This should be however re-examined in the context of modified gravity, case by case \cite{Wojnar:2018hal,Wojnar:2020fqi,sarmah2022stability}, or in a more general framework \cite{2015PhRvD..92d4046G,2022arXiv220612138K}.}, the microscopic variables must also be adjusted to the given theory of gravity. The reason for that is quite obvious: the definitions of the total entropy and the total particle number require a notion of the proper volume element, which is provided by the solution of the (modified) field equations, and by these means the model of gravity is taken into account in the thermodynamical variational principle. In the non-relativistic case, the modifications enter via the potential energy, that is, the gravitational potential, which resulted as a solution of the modified Poisson equation. In both cases, the thermodynamical quantities which carry the modifications and influence the equation of state, are the temperature and chemical potential, as already noticed in \cite{kulikov1995low}.

This work opens new research lines and questions such as, for instance: how significant are the effects introduced to microphysics by modified gravity? Let us notice that different models of modified gravity are effective models of quantum gravity proposals. Quantum gravity corrections to the equations of state are of Planck scale order \cite{verma2019effect,hamil2020effects}, and even so small modifications have an impact on the observable properties of stellar objects, such as for instance mass and radius \cite{moussa2015effect,harikumar2018compact}. Therefore, we should also expect a non-negligible effects in the case of Palatini gravity, even for small values of the theory parameters. Therefore, the data related to the critical masses of compact stars, such as maximal neutron stars masses or Chandrasekhar one, and of non-relativisitc objects, as minimum Main Sequence mass for example, could be used to constrain gravity. Further, we can also ask:
How should more general equations of state be modified to consistently describe a system at the statistical equilibrium? How nuclear reactions happening in the cores of stars should be modified in the light of those findings? How an extra degree of freedom, as for example scalar field or dynamical connection, influences the derivation of local and global variables, and, effectively, the equation of state? Is a stable thermodynamical equilibrium state also always dynamically stable in modified gravity, as it happens in GR? How does rotation affect our current considerations \cite{2022arXiv221211620C}? Some of those hypotheses are already investigated and the answers should appear elsewhere in the nearest future.

\noindent \textbf{Acknowledgement.} 
This work was supported by MICINN (Spain) {\it Ayuda Juan de la Cierva - incorporac\'ion} 2020 No. IJC2020-044751-I and by the EU through the European Regional Development Fund CoE program TK133 ``The Dark Side of the Universe.'' 

\bibliography{main.bib}

\end{document}